\pdfoutput=1
\documentclass[conference]{IEEEtran}
\IEEEoverridecommandlockouts
\usepackage{cite}
\usepackage{amsmath,amssymb,amsfonts}
\usepackage{algorithmic}
\usepackage{graphicx}
\usepackage{textcomp}
\usepackage{xcolor}

\def\BibTeX{{\rm B\kern-.05em{\sc i\kern-.025em b}\kern-.08em
    T\kern-.1667em\lower.7ex\hbox{E}\kern-.125emX}}

\usepackage{tikz}
\usepackage{pgfplots}\pgfplotsset{compat=1.9}
\usepackage{float}
\usepackage[caption = false]{subfig}
\PassOptionsToPackage{hyphens}{url}\usepackage{hyperref}
\newcommand{\norm}[1]{\lVert #1 \rVert}

\begin{document}

\title{Distributed Silhouette Algorithm: \\ Evaluating Clustering on Big Data
\thanks{Work done during Mater Thesis at Politecnico di Torino in 2015 under the supervision of Prof. Tania Cerquitelli and Prof. Paolo Garza and at Hortonwroks Inc.}
}

\author{\IEEEauthorblockN{Marco Gaido}
\IEEEauthorblockA{Trento, Italy \\
marcogaido91@gmail.com}
}

\maketitle

\begin{abstract}
In the big data era, the key feature that each algorithm needs to have
is the possibility of efficiently running in parallel in a distributed environment.
The popular Silhouette metric to evaluate the quality of a clustering, unfortunately,
does not have this property and has a quadratic computational complexity with
respect to the size of the input dataset. For this reason, its execution has been hindered
in big data scenarios, where clustering had to be evaluated otherwise.
To fill this gap, in this paper we introduce the first algorithm that computes
the Silhouette metric with linear complexity and can easily execute in parallel
in a distributed environment. Its implementation is freely available in the
Apache Spark ML library.
\end{abstract}

\begin{IEEEkeywords}
silhouette, clustering, Apache Spark
\end{IEEEkeywords}

\section{Introduction}

As the amount of data that is produced every day is huge and keeps increasing
the need for efficient solutions to process huge volumes of data has risen \cite{hadoop}.
These solutions solve the problem by parallelizing the work over different machines
that belong to a cluster. In this way, each machine processes a portion of the data
and the overall time required to perform the operations scales down.
Among the systems for distributed and parallel data processing, Apache Spark\footnote{\url{https://spark.apache.org/}.} is
nowadays the most widespread solution, as it allows for an easy and efficient execution
of the required transformations and the wide range of operations it supports.

In particular, Apache Spark also features the distributed execution of many supervised and unsupervised machine-learning algorithms,
which include clustering methods \cite{Rokach2005}, such as K-Means and Gaussian Mixture models.
The efficient implementation of these methods in a distributed environment is not complicated and their computational complexity
is linear with the size of the input dataset \cite{Velmurugan}.
However, the output of these clustering methods should also be evaluated with equal efficiency and this is non-trivial
as one of the most widespread clustering evaluation metrics, the Silhouette \cite{Rousseeuw}, has a quadratic computational
complexity with respect to the input size. This unfortunately holds true also
for the efficient implementation by \cite{efficient_sil}, which pre-compute (and cache) part of the operations 
and prevents its adoption in big data settings due to the excessive computational time required on huge datasets.

To overcome this limitation, and inspired by the idea of pre-computing part of the operations of \cite{efficient_sil},
in this work we describe the first algorithm that computes the Silhouette scores
with linear complexity with respect to the size of the dataset.
Our method requires a dedicated implementation for each distance measure and
currently it is defined (and described in this paper) for two distance measures:
\textit{i)} the squared euclidean distance, and \textit{ii)} the cosine distance.
The algorithm is easy to distribute on
different machines, being particularly suitable for parallel computation. In light of these appealing
characteristics, it has been implemented and contributed to the Apache Spark ML library
under the Apache 2.0 Licence and constitutes its current implementation.

\section{Background: The Silhouette Metric}

The Silhouette is a widespread metric used to evaluate the quality of a clustering operation.
Specifically, it is an unsupervised metric that measures how close the data in the same cluster are
in opposition to how separated they are from other clusters, in particular from the closest cluster to each
given datum (named ``neighbouring cluster'').

\paragraph{Definition} Formally, the Silhouette score $s_i$ for each datum $i$ is computed as the the difference between its
average distance to the other data in the same cluster $a_i$ and its average distance to the data in
the neighbouring cluster $b_i$, rescaled by the maximum of them to contain a value in the interval $[-1, 1]$:

\begin{equation}
s_{i} = \frac{b_{i}-a_{i}}{max\{a_{i},b_{i}\}}
\label{eq:silhouette}
\end{equation}

which can be rewritten as

\begin{equation}
\label{eq:silhouette2}
s_{i}=\left\{ \begin{tabular}{cc}
   $1-\frac{a_{i}}{b_{i}}$ & if $a_{i} \leq b_{i}$ \\
   $\frac{b_{i}}{a_{i}}-1$ & if $a_{i} > b_{i}$
\end{tabular} 
\right .
\end{equation}

The overall Silhouette score is then $S = \sum_{i=1}^{N}{s_{i}} / N$, i.e. the average of all $s_{i}, \forall i \in [1, ..., N]$, where $N$ is
the dataset size.
As such, the metric ranges from -1 to +1 and the highest it is,
the better the clustering is.

\paragraph{Computational Complexity}
As we do not know in advance which is the closest cluster to each datum,
the implementation of the Silhouette requires that, for each datum,
we compute its distance with all the points in the dataset and
average them by cluster. Once we have the average distance between one point
and all the clusters, $s_i$ can be easily computed with the equations above.
However, computing the distance between each datum and all the others is the dataset
has a computational complexity of $O(N^2 * D)$, where $D$ is the number of dimensions
in the given dataset, i.e. $X_i \in \mathbb{R}^D$. Indeed, the distance metric computation,
although depending on the actual distance considered, generally requires $O(D)$ operations
and we need to compute $O(N^2)$ distances.
As already discussed, this computational complexity leads to excessive computational costs
and time in a big data environment, where $N$ is a large number. In addition,
in a distributed environment, this computation either required that each machine hosts the whole dataset
to compute the distance with a datum -- which would cause an $O(N)$ memory footprint, causing OOM
issues for large $N$ -- or that the whole dataset is exchanged over the network between the machines (usually named \textit{workers})
at least for $W$ times, where $W$ is the number of workers used.
From this discussion, it is clear that the algorithm does not efficiently scale with the size
of the input dataset and with the number of workers, as required in big data clusters.

\section{Distributed Silhouette}
\label{sec:distrib_sil}

To avoid the above mentioned limitations of the Silhouette implementation,
we designed methods to compute it with a linear complexity with respect to the input size
and that allow for an efficient distribution of the workload across different workers.
Namely, the critic operation is the computation of the average distance between a datum $X_i$
and the points $C_j$ belonging to each cluster $\Gamma_k$ with $k \in [1, ..., K]$,
where $K$ is the number of clusters obtained from the execution of a clustering method.
As such, the focus of this work is the computation with linear complexity of such distances:

\begin{equation}
\label{eq:distances}
d(X_{i}, \Gamma_k) = \sum_{j=1}^{N_{\Gamma_k}}{d(X_{i}, C_{j})} / N_{\Gamma_k}
\end{equation}

for each $k$.
To do so, we designed specialized algorithms that strictly depend on the distance measure used.
In particular, in this work we describe the algorithms for two widespread distance measures:
the \textit{squared Euclidean distance} (\S\ref{subsec:squared_eucl}),
and the \textit{cosine distance} (\S\ref{subsec:cosine}).
Though, a similar approach may be used for other metrics as well.

\subsection{Squared Euclidean Distance}
\label{subsec:squared_eucl}

When using the squared Euclidean distance as distance measure,
the distance between one datum $X_{i}$ and a cluster $\Gamma_k$ can be rewritten as:

\begin{equation}
\label{eq:squaredExplosed}
\begin{aligned}
\sum\limits_{j=1}^{N_{\Gamma_k}} d(X_{i}, & C_{j})^2 = \sum\limits_{j=1}^{N_{\Gamma_k}} \Big( \sum\limits_{l=1}^D (x_{il}-c_{jl})^2  \Big) \\
& = \sum\limits_{j=1}^{N_{\Gamma_k}} \Big( \sum\limits_{l=1}^D x_{il}^2 + \sum\limits_{l=1}^D c_{jl}^2 -2\sum\limits_{l=1}^D x_{il}c_{jl}  \Big)  \\
& = \sum\limits_{j=1}^{N_{\Gamma_k}}\sum\limits_{l=1}^D x_{il}^2 + \sum\limits_{j=1}^{N_{\Gamma_k}} \sum\limits_{l=1}^D c_{jl}^2 - 2 \sum\limits_{j=1}^{N_{\Gamma_k}} \sum\limits_{l=1}^D x_{il}c_{jl} 
\end{aligned}
\end{equation}

where $D$ is the number of dimensions of the data in the dataset, $x_{il}$ is the $l$-th dimension of the $X_i$ vector, and $c_{jl}$ is the $l$-th
dimension of the $C_j$ vector belonging to the $\Gamma_k$ cluster, which contains $N_{\Gamma_k}$ elements.

Then, the first element of Eq. \ref{eq:squaredExplosed} can be rewritten as:

\begin{equation}
\sum\limits_{j=1}^{N_{\Gamma_k}}\sum\limits_{l=1}^D x_{il}^2 = N_{\Gamma_k} \xi_{X_i} \text{,  where  } \xi_{X_i} = \sum\limits_{l=1}^D x_{il}^2
\label{eq:csiX}
\end{equation}

where $\xi_{X_i} = \sum\limits_{l=1}^D x_{il}^2$ can be pre-computed independently and in parallel for each point $X_i$.

In addition, keeping in mind the definition of $\xi_{X_i}$, the second term of Eq. \ref{eq:squaredExplosed} can be rewritten as:

\begin{equation}
\label{eq:psi}
\sum\limits_{j=1}^{N_{\Gamma_k}} \sum\limits_{l=1}^D c_{jl}^2 = \sum\limits_{j=1}^{N_{\Gamma_k}} \xi_{C_{j}} = \Psi_{\Gamma_k}
\end{equation}

which can be pre-computed for each cluster with a single pass over the whole dataset, i.e. with linear complexity with respect
to the size of the dataset.\footnote{In the implementation, each worker can independently compute the cluster-wise sums of its data,
which are then collected on a node which aggregates them in the overall sums. We can notice that the efficiency of this algorithm, hence,
depends on the number of clusters, which in practice is a fairly small number and much lower than the dataset size.}

Lastly, the last element of Eq. \ref{eq:squaredExplosed} can be rewritten as:

\begin{equation}
\label{eq:ypsilon1}
\sum\limits_{j=1}^{N_{\Gamma_k}} \sum\limits_{l=1}^D x_{il}c_{jl} =  \sum\limits_{l=1}^D \Big(\sum\limits_{j=1}^{N_{\Gamma_k}} c_{jl} \Big) x_{il} 
\end{equation}

and, by defining a vector $Y_{\Gamma_k}$ so that:

\begin{equation}
\label{eq:ypsilon}
Y_{\Gamma_{k}l} = \sum\limits_{j=1}^{N_{\Gamma_k}} c_{jl} \forall l \in [1, ..., D]
\end{equation}

we obtain:

\begin{equation}
\label{eq:ypsilon2}
\sum\limits_{l=1}^D \Big(\sum\limits_{j=1}^{N_{\Gamma_k}} c_{jl} \Big) x_{il}  = \sum\limits_{l=1}^D Y_{\Gamma_{k}l} x_{il}
\end{equation}

where the vectors $Y_{\Gamma_{k}}$ can be pre-computed for each cluster with a single pass over the whole dataset, similarly to $\Psi_{\Gamma_k}$.\footnote{In the implementation, $Y$ and $\Psi$ are jointly computed with a single pass over the whole dataset.}

As such, by integrating Eq. \ref{eq:csiX}, Eq. \ref{eq:psi} and Eq. \ref{eq:ypsilon2} into Eq. \ref{eq:squaredExplosed}, we can rewrite Eq. \ref{eq:squaredExplosed} as:

\begin{equation}
\label{eq:rootSquaredSilNumerator}
N_{\Gamma_k} \xi_{X_i} + \Psi_{\Gamma_k} - 2 \sum\limits_{l=1}^D Y_{\Gamma_{k}l} x_{il}.
\end{equation}

With this formula, the average distance of the datum $X_i$ from cluster $\Gamma_k$ (Eq. \ref{eq:distances}) becomes:

\begin{equation}
\label{eq:squaredRootSilhouetteSingle}
\begin{aligned}
\frac{\sum\limits_{j=1}^{N_{\Gamma_k}} d(X_i, C_{j})^2}{N_{\Gamma_k}} & = \\
& = \frac{N_{\Gamma_k} \xi_{X_i} + \Psi_{\Gamma_k} - 2 \sum\limits_{l=1}^D Y_{\Gamma_{k}l} x_{il}}{N_{\Gamma_k}} \\
& = \xi_{X_i} + \frac{\Psi_{\Gamma_k}}{N_{\Gamma_k}} - 2 \frac{\sum\limits_{l=1}^D Y_{\Gamma_{k}l} x_{il}}{N_{\Gamma_k}} 
\end{aligned}
\end{equation}

In this way, the distance between each element $X_i$ and each cluster $\Gamma_k$
does not require the computation of the distance between $X_i$ and all the other data in the dataset. Indeed,
each $X_i$ can be processed independently, as
it is enough to pre-compute the constant $\xi_{X_i}$ for each point $X_i$ and
the constants $\Psi_{\Gamma_k}$ and $N_{\Gamma_k}$ and the vector $Y_{\Gamma_{k}}$ for each cluster $\Gamma_k$.
In the Apache Spark implementation, the pre-computed values for the clusters are distributed among the worker nodes via broadcasted variables,
because we can assume that the clusters are limited in number and anyway they are much fewer than the points.

The main strengths of this algorithm are the low computational complexity and the intrinsic parallelism.
As we have seen, $\Psi_{\Gamma_k}$, $N_{\Gamma_k}$ and the vector $Y_{\Gamma_{k}}$ can be pre-computed with a computational complexity
that is $O(D * N / W)$.
After that, every point can be analyzed independently of the others.
Specifically, for every point we need to compute the average distance to all the clusters.
Since, Eq. \ref{eq:squaredRootSilhouetteSingle} requires $O(D)$ operations,
this phase has a computational complexity of $O(C * D * N / W)$ where $C$ is the number of clusters (which we assume quite low).
Lastly, each score $s_i$ can be computed with Eq. \ref{eq:silhouette2}, and the $s_i$ scores are averaged.
This average has a computational complexity of $O(N/W)$.
All in all, we can conclude that the computational complexity of the algorithm is $O(C * D * N / W)$.
As in big data settings it is reasonable to assume that $N >> C$ and $N >> D$, this is $O(N/W)$
that means that the algorithm scales linearly with the size of the input dataset and that
the time required to compute the metric reduces linearly with the number of workers used.
This is an ideal condition in big data clusters as it enforces that the size of a dataset
can grow indefinitely without increasing the time required to perform the computation by
scaling with the same growth factor the number of worker nodes.

\subsection{Cosine Distance}
\label{subsec:cosine}

To define the metric with the cosine distance, we use a similar approach.
The cosine distance is defined as $1-cs$, where $cs$ is the cosine similarity, which is:

\begin{equation}
\label{eq:cosien_sim}
cs(X, Y) = \frac{\sum\limits_{l=1}^D x_{l} y_{l}}{\lVert X \rVert \lVert Y \rVert}.
\end{equation}

Hence, the average distance between a datum $X_i$ and the data $C_{j}$ of a cluster  $\Gamma_k$ is: 

\begin{equation}
\label{eq:firstCosineSilPax}
\frac{\sum\limits_{j=1}^{N_{\Gamma_k}} d(X_{i}, C_{j} ) }{N_{\Gamma_k}}  = 
\frac{\sum\limits_{j=1}^{N_{\Gamma_k}} \Bigg(1-\frac{ \sum\limits_{l=1}^D x_{il}c_{jl}}{\norm{X_i}\norm{C_{j}}} \Bigg)}{N_{\Gamma_k}}.
\end{equation}

The numerator can be rewritten as:

\begin{equation}
\label{eq:secCosineSilPax}
\begin{aligned}
    \sum\limits_{j=1}^{N_{\Gamma_k}} 1 & - \sum\limits_{j=1}^{N_{\Gamma_k}} \sum\limits_{l=1}^D \frac{x_{il}}{\norm{X_i}}\frac{c_{jl}}{\norm{C_{j}}} \\
    & = N_{\Gamma_k} -  \sum\limits_{l=1}^D \frac{x_{il}}{\norm{X_i}} \Bigg(\sum\limits_{j=1}^{N_{\Gamma_k}}\frac{c_{jl}}{\norm{C_{j}}}\Bigg).
\end{aligned}
\end{equation}

Now, analogously to the squared Euclidean case, we can define the vectors

\begin{equation}
\label{eq:cosineCsi}
\xi_{X_i} : \xi_{X_{i}l} = \frac{x_{il}}{\norm{X}} \forall l \in [1, ..., D]
\end{equation}

which can be pre-computed for each datum and

\begin{equation}
\label{eq:omega}
\Omega_{\Gamma_k} : \Omega_{\Gamma_{k}l} = \sum\limits_{j=1}^{N_{\Gamma_k}} \xi_{C_{j}l} \forall l \in [1, ..., D]
\end{equation}

which can be pre-computed for each cluster. Eq. \ref{eq:secCosineSilPax} hence becomes:

\begin{equation}
\label{eq:cosineNumerator}
N_{\Gamma_k} -  \sum\limits_{l=1}^D \xi_{X_{i}l} \Omega_{\Gamma_{k}l}.
\end{equation}

Therefore, Eq.  \ref{eq:firstCosineSilPax} can be computed with

\begin{equation}
\label{eq:cosineSilhouette}
1 -  \frac{\sum\limits_{l=1}^D \xi_{X_{i}l} \Omega_{\Gamma_{k}l}}{N_{\Gamma_k}}
\end{equation}

which can be computed for each $X_i$ without comparing it to all the data in other clusters,
but using only the above-defined pre-computed vectors. 
Once obtained the average distance
of each $X_{i}$ for all the clusters, its Silhouette score $s_{i}$ is computed with Eq. \ref{eq:silhouette2}
and the Silhouette scores are averaged for all the elements in the dataset.
As can be inferred from its definition, all the considerations regarding the computational costs done in the
previous subsection for the squared Euclidean distance apply to this case as well.

\section{Experiments}

\pgfplotstableread[row sep=\\,col sep=&]{
N & std & proposed \\
440 & 1.996 & 2.21 \\
6435 & 124.667 & 4.279 \\
9901 & 159.941 & 4.735 \\
14030 & 1936.031 & 14.574 \\
20000 & 699.314 & 4.796 \\
137292 & 14392.835 & 20.97 \\
149009 & 20498.115 & 10.606 \\
149724 & 16644.212 & 23.358 \\
159741 & 23035.701 & 12.105 \\
161050 & 23401.273 & 17.585 \\
164879 & 29824.76 & 18.567 \\
166198 & 29426.827 & 16.815 \\
170540 & 32987.117 & 15.046 \\
}\fixedsegmlen

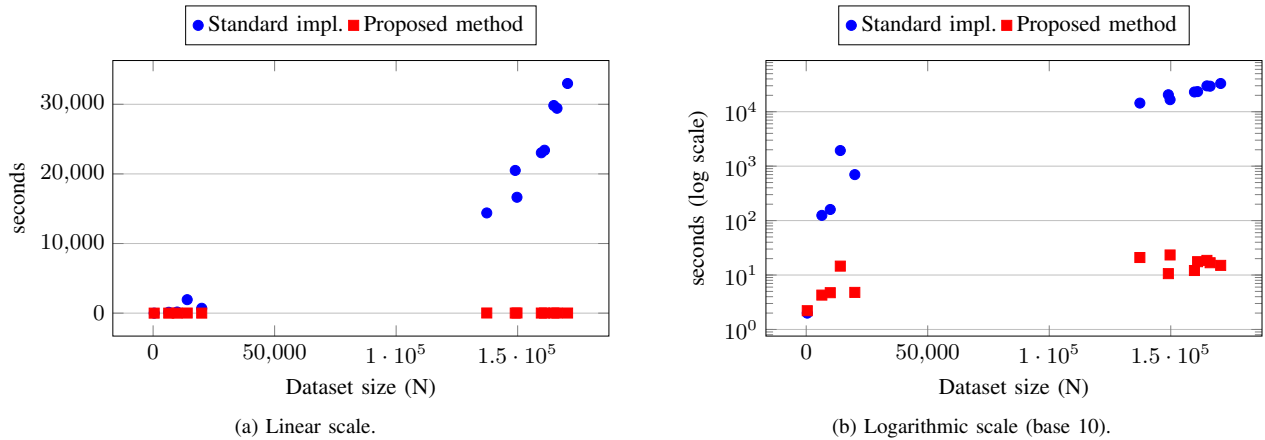
\begin{figure*}[t]
\centering
    \subfloat[Linear scale.]{\label{fig:std}
\begin{tikzpicture}[scale=0.95]
    \small
    \begin{axis}[
            ymajorgrids=true,
            width=.47\textwidth,
            height=.3\textwidth,
            legend style={at={(0.5,1.2)},
            anchor=north,legend columns=-1},
            xtick=data,
            xtick={},
            scaled y ticks=false,
            scaled x ticks=false,
            only marks,
            ylabel={seconds},
            xlabel={Dataset size (N)}
        ]
        \addplot[color=blue, mark=*] table[x=N,y=std]{\fixedsegmlen};
        \addplot[color=red, mark=square*] table[x=N,y=proposed]{\fixedsegmlen};
        \legend{Standard impl., Proposed method}
    \end{axis}
\end{tikzpicture}
}
\qquad
    \subfloat[Logarithmic scale (base 10).]{\label{fig:log}
\begin{tikzpicture}[scale=0.95]
    \small
    \begin{axis}[
            ymajorgrids=true,
            width=.47\textwidth,
            height=.3\textwidth,
            legend style={at={(0.5,1.2)},
            anchor=north,legend columns=-1},
            xtick=data,
            xtick={},
            scaled y ticks=false,
            scaled x ticks=false,
            only marks,
            log basis y={10},
            ymode=log,
            ylabel={seconds (log scale)},
            xlabel={Dataset size (N)}
        ]
        \addplot[color=blue, mark=*] table[x=N,y=std]{\fixedsegmlen};
        \addplot[color=red, mark=square*] table[x=N,y=proposed]{\fixedsegmlen};
        \legend{Standard impl., Proposed method}

    \end{axis}
\end{tikzpicture}
}
\caption{\label{fig:times} Runtime required (in seconds) to compute the Silhouette score (with squared Euclidean distance) with the standard implementation and our approach.}
\end{figure*}

To showcase the benefits of the proposed algorithm, we compare the runtime
required by the standard Silhouette implementation and the method proposed in this paper
with the squared Euclidean distance as a dissimilarity measure.
All the experiments have been executed with a single thread on a  MacBook Pro
with 2,8 GHz Intel Core i7 and 8 GB of RAM 1600 MHz DDR3.
In this condition, we do not exploit the ability of our method to parallelize
over multiple workers. So in a big data scenario, the difference would be even larger 
or the standard implementation would not be an option in case of very large datasets.
We used a proprietary dataset with 129 features and observed the computational cost
by increasing the dataset size.
The results are reported in Fig. \ref{fig:times}.

First, we can notice that the quadratic complexity of the standard implementation emerges clearly in Fig. \ref{fig:std}.
In addition, when we reach the 100,000 of dataset size, its runtime explodes in comparison with our proposed method
with a difference of 3 orders of magnitude, as we can see from Fig. \ref{fig:log}. With the proposed method, the runtime
reaches at maximum 23 seconds and it is not reached at the maximum dataset size: this happens because with different
sizes we also have a different number of clusters in this experiment and, as previously seen, the number of clusters
plays an important role in the computational cost of our method.

\section{Conclusions}

With the goal of enabling the execution and proper evaluation of clustering algorithms in a big data environment,
in this work we describe the first method to compute the Silhouette score with a linear computational
complexity with respect to the input dataset size and with the possibility of being executed in parallel over different machines.
Our scaling experiment, although performed on relatively small datasets ($\sim$150,000 data), showed the
great benefits of the proposed algorithm that are even larger in a real distributed environment.
The implementation of our method has been contributed to the Apache Spark ML library with Apache 2.0 Licence
and is also available at \url{https://gitlab.com/mark91/SparkClusteringEvaluationMetrics/-/tree/master/src/main/scala/org/apache/spark/ml/evaluation}
under the same license.

\section{Limitations}

As discussed in \S\ref{sec:distrib_sil}, the proposed approach does not provide a generic
algorithm that generalizes over any distance measure. On the contrary, it requires a dedicated
implementation for each distance measure. Currently, the algorithms and implementations have been
defined only for the squared Euclidean distance and the cosine distance, but similar definitions
may be possible also for other distance measures.
Unfortunately, it is hard to apply the same approach to the Euclidean distance, as the square root
operator hinders the possibility of aggregating cluster-level statistics to be used for the final formula.
While approximations are possible, e.g. by taking the square root of the element-cluster average distances
the exact computation of the Silhouette with Euclidean distance with this method is not possible.

%\section*{References}

%\bibliography{biblio}
%\bibliographystyle{natbib}

\end{document}